\documentclass[reprint,amsmath,amssymb,pra,aps,showpacs,superscriptaddress,twocolumn,nofootinbib,nobibnotes,bibnotes]{revtex4-2}

\usepackage{amsfonts,amssymb,amsthm,graphics,graphicx,epsfig,bbm}
\usepackage[colorlinks=true,citecolor=blue,linkcolor=blue,urlcolor=blue]{hyperref}
\usepackage[usenames]{color}
\usepackage{graphicx}
\usepackage{subfigure}
\usepackage{dcolumn}
\usepackage{bm}
\usepackage{times}
\usepackage{epstopdf}
\usepackage{amstext}
\usepackage{latexsym}
\usepackage{hyperref}
\usepackage{psfrag}
\usepackage{soul,xcolor}
\usepackage[normalem]{ulem}
\usepackage{dsfont}
\usepackage{txfonts}
\usepackage{float}
\usepackage{titlesec}

\titlespacing\section{0pt}{4pt plus 4pt minus 2pt}{0pt plus 2pt minus 2pt}
\titlespacing\subsection{0pt}{4pt plus 4pt minus 2pt}{0pt plus 2pt minus 2pt}

\renewcommand{\Re}{\mathrm{Re}}

\newcommand{\ket}[1]{\vert #1 \rangle}
\newcommand{\bra}[1]{\langle #1 \vert}

\newcommand{\abs}[1]{| #1 |}

\begin{document}

\preprint{APS/123-QED}

\title{Quantum Memristors with Quantum Computers}

\author{Y.-M. Guo}
\affiliation{International Center of Quantum Artificial Intelligence for Science and Technology (QuArtist) \\  and Physics Department, Shanghai University, 200444 Shanghai, China}

\author{F. Albarr\'an-Arriagada}
 \email{pancho.albarran@gmail.com}
\affiliation{International Center of Quantum Artificial Intelligence for Science and Technology (QuArtist) \\  and Physics Department, Shanghai University, 200444 Shanghai, China}

\author{H. Alaeian}
\affiliation{Elmore Family School of Electrical and Computer Engineering, Department of Physics and Astronomy, Purdue Quantum Science and Engineering Institute, Purdue University, West Lafayette, IN 47907, USA}

\author{E. Solano}
 \email{enr.solano@gmail.com}
\affiliation{International Center of Quantum Artificial Intelligence for Science and Technology (QuArtist) \\  and Physics Department, Shanghai University, 200444 Shanghai, China}
\affiliation{IKERBASQUE, Basque Foundation for Science, Plaza Euskadi 5, 48009 Bilbao, Spain}
\affiliation{Kipu Quantum, Kurwenalstrasse 1, 80804 Munich, Germany}

\author{G. Alvarado Barrios}
 \email{phys.gabriel@gmail.com}
\affiliation{International Center of Quantum Artificial Intelligence for Science and Technology (QuArtist) \\  and Physics Department, Shanghai University, 200444 Shanghai, China}

\date{\today}

\begin{abstract}
We propose the encoding of memristive quantum dynamics on a digital quantum computer. Using a set of auxiliary qubits, we simulate an effective non-Markovian environment inspired by a collisional model, reproducing memristive features between expectation values of different operators in a single qubit. We numerically test our proposal in an IBM quantum simulator with 32 qubits, obtaining the pinched hysteresis curve that is characteristic of a quantum memristor. Furthermore, we extend our method to the case of two coupled quantum memristors, opening the door to the study of neuromorphic quantum computing in the NISQ era.
\end{abstract}

\maketitle

\section{Introduction}~\label{sec:1}

The memristor was theoretically proposed in 1971 by Leon Chua~\cite{1971Chua,1976Chua} as a two-terminal passive circuit element relating the charge and the flux. As a consequence of this relation, the current across the device depends on the history of charges that have passed through it. Specifically, a voltage-controlled memristor has the circuit variable relations
\begin{eqnarray}~\label{memristor_characterstic_eq}
    &I(t) = M(q)V(t)\, ,\nonumber\\
    &q = \int I(t) dt\, .
\end{eqnarray}
The memductance $M(q)$ depends on the internal state given by the charge $q$, which depends on the history of the current flowing through the device. With this state-dependent Ohm's law, the independent variable $V(t)$ and dependent variable $I(t)$ constitute the input and output of a memristor, respectively. 
Despite its apparent simplicity, the first experimental realization of a memristor was only achieved in 2008 by Hewlett-Packard labs~\cite{2007WaserR,2008Strukov} and, since then, the field of memristive devices has developed substantially~\cite{Thomas2013JPD,Marani2015arXiv,Vourkas2016IEEE,Li2018JPD}. These devices possess memory effects and a nonlinear current-voltage relationship. In this sense, they are one of the leading candidates for the implementation of neuromorphic computers~\cite{Wang2018NM, Lin2020NE, Yao2020Nat} that  may overcome the von Neumann bottleneck~\cite{Upadhyay2019AMT}.  

Given the promising prospects of a memristor as a fundamental element for neuromorphic classical computing, it is natural to ask whether it is possible to design a quantum version of this element and explore its properties. Within the last years, several theoretical efforts have investigated this question~\cite{2014Sebastiano,2016Pfeiffer,2017Salmilehto,2018Sanz M,2020Gonzalez}, and some quantum memristive devices have been already implemented in photonics platforms~\cite{Spagnolo2021arXiv, Gao2020arXiv, Zhou2020arXiv}. Nevertheless, the implementation of small quantum memristive networks has eluded experimental efforts due to the difficulty involved in coupling quantum memristive devices.

On the other hand, universal quantum computers offer the possibility of simulating complex system dynamics in a digital way~\cite{Georgescu2014RMP}. Therefore, the existing technologies allow the experimental realization of small and noisy quantum devices as good candidates for quantum simulations of complex quantum dynamics with unitary gates. However, quantum memristive systems involve non-unitary dynamics of non-Markovian nature. The latter can be obtained from the unitary dynamics of a larger system, tracing some degrees of freedom associated with auxiliary subsystems commonly called environment. This means we can obtain effective families of non-unitary and non-Markovian dynamics in quantum computers by considering a subset of quantum processor qubits~\cite{GarciaPerez2020NPJQI, HeadMarsden2021PRR}. 

In this manuscript, we simulate the dynamics of a quantum memristor on a quantum computer by digitally implementing its open system dynamics. We test our protocol for a single two-level quantum memristor in an IBM quantum simulator with 32 qubits. In addition, we study coupled quantum memristors with a variety of possible interactions. This work showcases how quantum computers can be used as a testbed for studying individual and coupled quantum memristors, an important step towards thorough studies of memristor-based neuromorphic quantum computing.

This paper is organized as follows: in section~\ref{sec:2}, we present the model of a two-level quantum memristor. In section~\ref{sec:3}, we describe the developed protocols for the digital simulation of the single and coupled quantum memristors, in general, and demonstrate the feasibility of the algorithm by showing several examples. Section~\ref{sec:4} concludes this work and presents further possible developments.

\vspace{0.5cm}

\section{Two-level quantum memristor}~\label{sec:2}

Let us consider the theoretical proposal of the conductance-asymmetric SQUID quantum memristor~\cite{2017Salmilehto}. This system is described by a quantum harmonic oscillator with a time-dependent quasiparticle decay rate $\Gamma(t)$ at zero temperature. The dynamics of the density matrix $\hat{\rho}$ is given by the following master equation
\begin{eqnarray}
\label{mastereq1}
\frac{\partial}{\partial t}\hat \rho {\rm{ = }} - \frac{i}{\hbar }[{\hat H},\hat \rho ]{\rm{ + }}{\Gamma }(t) \left({\hat a\hat \rho {{\hat a}^\dag } - \frac{1}{2}{{\hat a}^\dag }\hat a\hat \rho  - \frac{1}{2}\hat \rho {{\hat a}^\dag }\hat a} \right),
\end{eqnarray}
where $\hat{H}=\hbar\omega(\hat{a}^{\dagger}a + 1/2)$ and $\Gamma(t)>0$. 

As we are not considering external energy source, the system loses its energy continuously and the number of excitations in the oscillator, $\langle\hat{a}^{\dagger} \hat{a}\rangle$, decays in time. The current and voltage of this quantum memristor can be written as~\cite{2017Salmilehto}
\begin{eqnarray}
    \label{mem-eqs}
	\langle {\hat V} \rangle  &=& -\frac{-e}{2g_0C_d}\left\langle i(\hat{a}^{\dagger}-\hat{a}) \right\rangle,  \nonumber\\
	\langle \hat {I} \rangle  &=& \frac{e}{2g_0}\partial _t\left\langle i(\hat{a}^{\dagger}-\hat{a}) \right\rangle + \frac{4eg_0E_L}{\hbar}\left\langle (\hat{a}^{\dagger}+\hat{a}) \right\rangle.
\end{eqnarray}
Using Eq.~(\ref{mastereq1}), for the time derivative of the expectation value in Eq.~(\ref{mem-eqs}), it can be shown that the voltage is related to the current by the following expression
\begin{equation}
    \label{quantum_memristor_eq}
    \langle \hat I \rangle = G(t) \langle {\hat V} \rangle,
\end{equation}
where $G(t) = -e\Gamma(t) $ is the memductance. This equation has the same form of Eq.~(\ref{memristor_characterstic_eq}), which characterizes a quantum memristor. In consequence, when the input $\langle {\hat V} \rangle$ has a sinusoidal time dependence, then the current $\langle \hat I \rangle$ as a function of voltage $\langle {\hat V} \rangle$ traverses the typical pinched hysteresis loop of a quantum memristor.

\begin{figure}[t]
	\centering
	\includegraphics[width=1.05\linewidth]{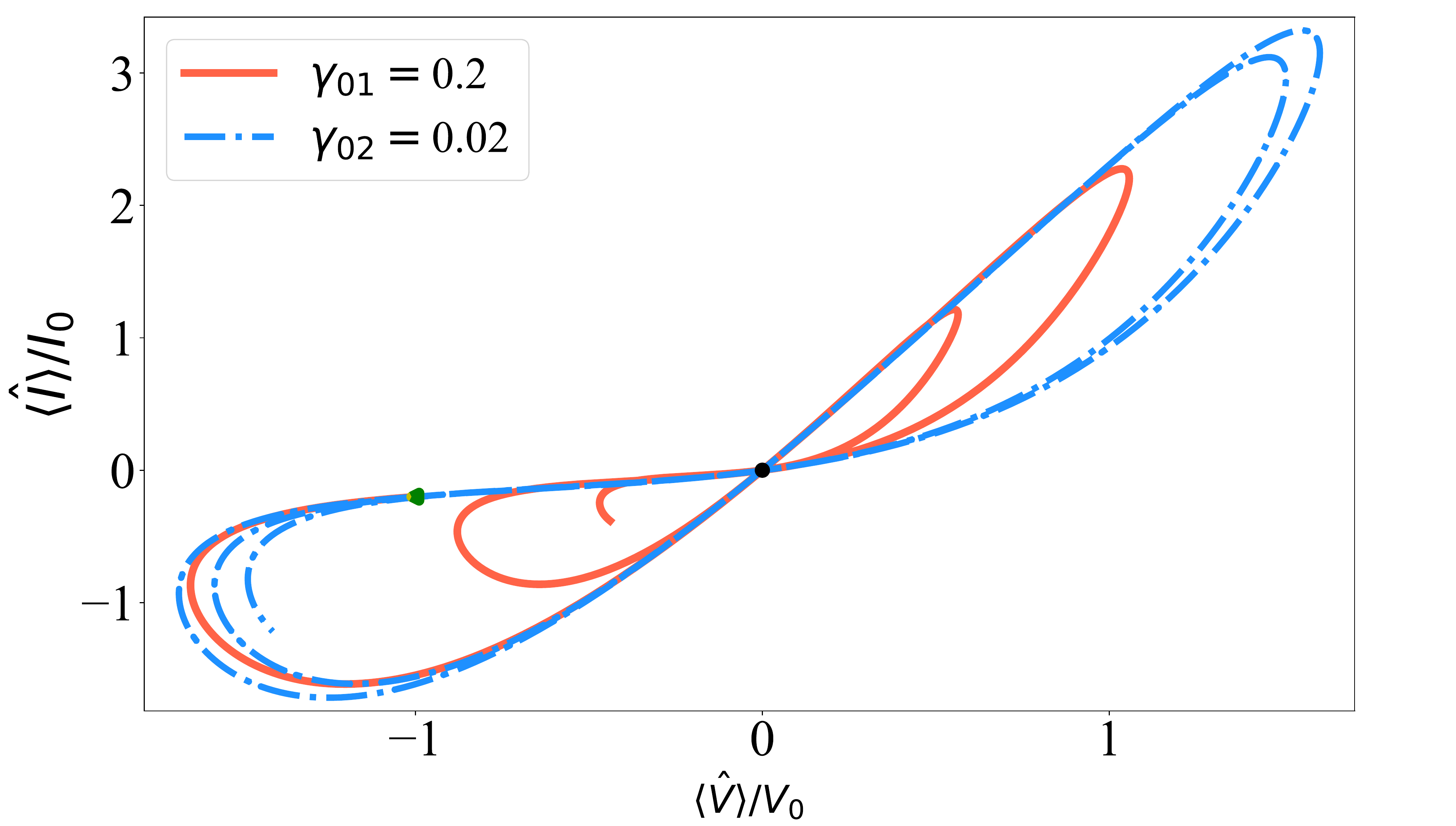}
	\caption{I-V characteristics of the two-level quantum memristors for two cases. For both quantum memristors, $m=1$, $\hbar=1$, and $\omega=1$, with the same initial state ${\vert \psi_{0} \rangle = \cos(\pi/8)\ket{e} + \sin(\pi/8)e^{(i\pi/5)}\ket{g}}$ and different decay rate $\gamma(t)$, as shown in Eq.~(\ref{gamma(t)}), $\gamma_{01}=0.2$ (solid red line) , $\gamma_{02}=0.02$ (dashed-dotted blue line). Expectation values are normalized to the initial values shown by the green dot, i.e. $V_0=\abs{\langle {\hat V} \rangle}_{t=0}$ and $I_0=\abs{\langle {\hat I} \rangle}_{t=0}$. The I-V curve in both cases show a pinched hysteresis loop, denoted by the black dot in the figure.}
	\label{Fig01}
\end{figure}

If we consider only one excitation in the initial state, at zero temperature, the dynamics can be approximated with a two-level system, where the Hamiltonian reads
\begin{equation}
	\hat{H_2}=\hbar\omega\left(\hat\sigma_+ \sigma_- + \frac{1}{2}\right) = \frac{1}{2}\hbar\omega \left(\hat\sigma_z+2\right),
\end{equation}
and the master equation becomes
\begin{eqnarray}
    \label{mastereq2}
	\frac{\partial}{\partial t}\hat \rho_2 {\rm{ = }} - \frac{i}{\hbar }[\hat{H_2},\hat {\rho_2} ] + \Gamma(t)\left(\hat{\sigma}_-\hat{\rho_2}\hat{\sigma}_+ - \frac{1}{2}\{\hat{\sigma}_+\hat{\sigma}_-,\hat{\rho_2}\}\right),
\end{eqnarray}
where $\hat{\sigma}_+,\hat{\sigma}_-$, and $\sigma_z$ are the Pauli raising, lowering, and z operators, respectively. 
We use a time-dependent decay rate given by Ref.~\cite{2017Salmilehto}, which reads
\begin{equation}\label{gamma(t)}
	\Gamma(t)=\gamma_0(1-\sin[\cos(\omega t)]),
\end{equation}
where $\gamma_{0}$ is a constant associated with the decay strength. Following Eq.~(\ref{mem-eqs}), the memristive variables can be written as
\begin{eqnarray}
    \label{2level_memristive_equation}
	&\langle {\hat V_2} \rangle  = -\frac{1}{2}\sqrt{\frac{m\hbar\omega}{2}}\langle \hat\sigma_y \rangle \, , \nonumber \\
	& \langle \hat I_2 \rangle  = \sqrt{\frac{m\hbar\omega}{2}}\frac{d}{dt}\langle \hat\sigma_y \rangle - \sqrt{\frac{m\omega}{2\hbar}}\langle \hat\sigma_x \rangle.
\end{eqnarray}
Now, the memristive equation now reads
\begin{equation}
    \langle \hat I_2 \rangle = \Gamma(t) \langle {\hat V_2} \rangle.
\end{equation}
As an example, we numerically solve Eq.~(\ref{mastereq2}) for different values of $\gamma_0$. In Fig.~{\ref{Fig01}}, we plot the corresponding current-voltage (I-V) curves showing the characteristic pinched hysteresis loop. In that figure, the red curve shrinks faster that the blue curve due to its larger decay rate. In both case, the system is initialized in a pure state as $\ket{\psi_{0}} = \cos(\pi/8)\ket{e} + \sin(\pi/8)e^{i\pi/5}\ket{g}$.

In next section, we show how the two-level quantum memristive dynamics can be simulated on a digital quantum computer using one qubit as the memristive system (system qubit) and a set of auxiliary qubits as a non-Markovian reservoir.

\vspace{0.5cm}

\section{Quantum circuits for memristive dynamics}~\label{sec:3}

\subsection{Single memristive dynamics}~\label{subsec:3A}

We describe how to map the memristive dynamics of Eq.~(\ref{mastereq2}) on a quantum circuit, where each digital step corresponds to a set of operations evolving the system qubit-state from time $t_{i}$ to $t_{i+1}$. The whole time evolution, therefore, can be simulated by the repeated application of these digital steps. This method can also be extended to incorporate interactions between memristors as we show later. 

We start by writing the master equation of Eq.~(\ref{mastereq2}) in the interaction picture
\begin{eqnarray}\label{two-level master equation In}
	\partial _t\hat{\rho}_I = \gamma(t)\left(\hat{\sigma}_-\hat{\rho}_I\hat{\sigma}_+ - \frac{1}{2}\{\hat{\sigma}_+\hat{\sigma}_-,\hat{\rho}_I\}\right),
\end{eqnarray}
where $\hat{\rho}_I(t)=e^{\frac{it}{\hbar}\hat H}\hat{\rho_2}(t)e^{-\frac{it}{\hbar}\hat H}$, and $\hat{H} \equiv \hat{H}_{2}$. At zero temperature, this system has an exact solution as (see Ref.~\cite{GarciaPerez2020NPJQI})
\begin{equation}
	\hat{\rho}_I(t) =  \begin{pmatrix}
		|c_1(t)|^2 & c_0^*c_1(t) \\
		c_0c_1(t)^* & 1-|c_1(t)|^2
	\end{pmatrix},
\end{equation}
where
\begin{eqnarray}
	\gamma (t) =  - 2\Re\left\{ \frac{\dot {c}_1(t)}{c_1(t)} \right\},
	\label{eq_for_c1}
\end{eqnarray}
and $c_0$ is determined by the initial state. Without loss of generality, we assume $c_1(t)$ to be a real number obtaining the following density matrix 
\begin{equation}\label{final exact solution}
	\hat{\rho}_I(t) =  \begin{pmatrix}
		c_1(t)^2 & c_0^*c_1(t) \\
		c_0c_1(t) & 1-c_1(t)^2
	\end{pmatrix}.
\end{equation}
Finally, Eq.~(\ref{eq_for_c1}) can be used to solve for $c_1(t)$ as
\begin{eqnarray}\label{c1(t)}
	{c_1}(t) = {c_1}(0)e^{\kappa(t)} \, , \nonumber\\
	\kappa(t) = \frac{{ - \int_0^t {\gamma (t')dt'} }}{2}.
\end{eqnarray}
If the initial state is a pure state of the form ${\vert \psi_{0} \rangle = \cos(a)\ket{e} + \sin(a)e^{ib}\ket{g}}$, the density matrix at time $t$ has the following form
\begin{equation}\label{final exact solution 2}
	\hat{\rho}_I(t) =  \begin{pmatrix}
		(\cos{a}{e^{\kappa(t)}})^2 & \cos{a}\sin{a}e^{ib}{e^{\kappa(t)}} \\
		\cos{a}\sin{a}e^{-ib}{e^{\kappa(t)}} & 1-({\cos{a}{e^{\kappa(t)}}})^2
	\end{pmatrix},
\end{equation}
where $a\in [0,\pi/2]$ and $b\in [0,2\pi]$.
\begin{figure}[t]
	\centering
	\includegraphics[width=1\linewidth]{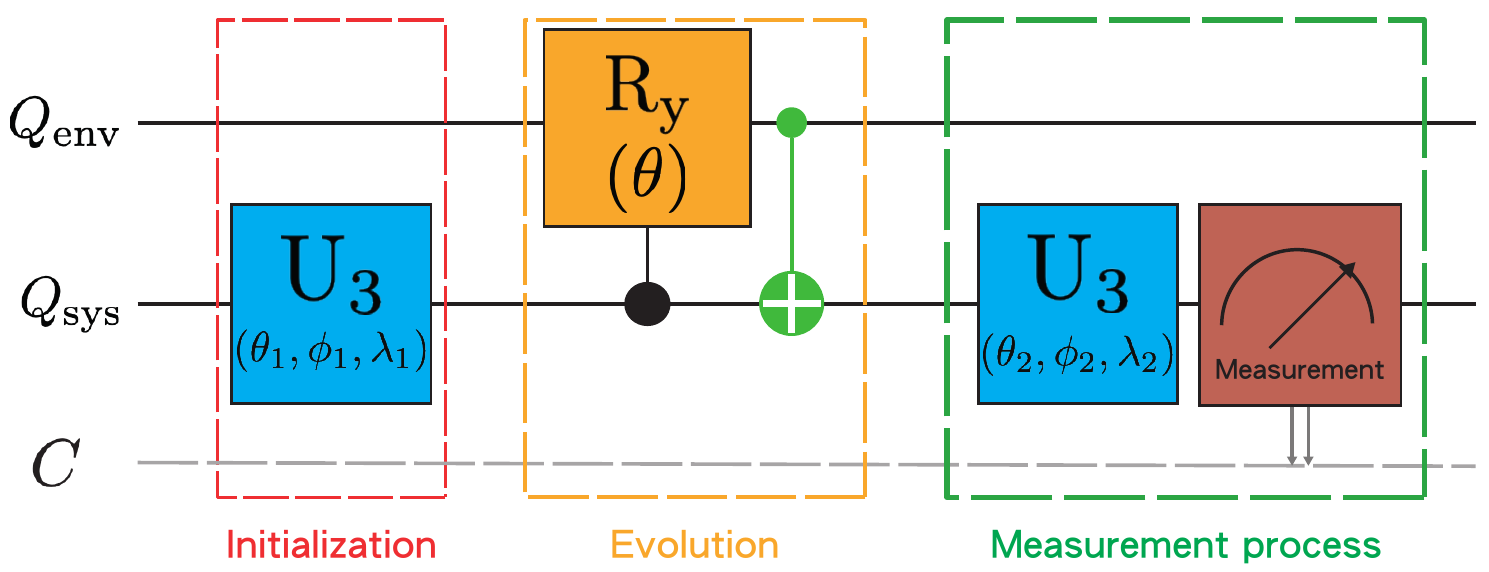}
	\caption{The proposed circuit for a single time-step quantum memristor simulation with three basic steps of initialization, evolution, and measurement process. The circuit consists of single-qubit gates, i.e. unitary gate (U3-in blue) and two-qubit gates, i.e. control-rotation-Y (in yellow) and CNOT gate (in green). The last step is a measurement process.}
	\label{Fig02}
\end{figure}

We can also write the dynamical map defined by the  master equation in~Eq.~(\ref{two-level master equation In}), corresponding to an amplitude damping mechanism, using its Krauss operators as~\cite{Nielsen2010Book}
\begin{eqnarray}
    \label{Amplitude_damping_map_1}
    &\hat{\rho}_I(t) = \varepsilon_{t,0}(\hat{\rho}_I(0)) = \hat{E}_0^{(0, t)} \hat{\rho}_I(0) (\hat{E}_0^{(0, t)})^{\dagger} + \hat{E}_1^{(0, t)}\hat{\rho}_I(0) (\hat{E}_1^{(0, t)})^{\dagger}, \nonumber\\ 
    &\hat{E}_0^{(0, t)} = \begin{pmatrix}
		e^{\kappa(t)} & 0 \\ 
		0  & 1
	\end{pmatrix} ~ , ~
	\hat{E}_1^{(0, t)} = \begin{pmatrix}
		0 & 0 \\
		\sqrt{1-e^{2\kappa(t)}}  & 0
	\end{pmatrix}.
\end{eqnarray}
Since the map of Eq.~(\ref{Amplitude_damping_map_1}) give us the density matrix in the interaction picture, we must transform it back to the Schr\"{o}dinger picture using  ${\hat{\rho}_2(t)=e^{-\frac{it}{\hbar}\hat H}\hat{\rho}_I(t)e^{\frac{it}{\hbar}\hat H}}$, to calculate the expectation values of the two-level quantum memristor.

To mimic the action of the operators $\hat{E}_0$ and $\hat{E}_1$, we consider the gate circuit shown in Fig.~\ref{Fig02}. This gate-based circuit is split into three basic steps. First, the initialization, where we prepare the qubit $Q_{\textrm{sys}}$ in its initial state. Second, the evolution step, where we entangle the system and environment qubits, using a controlled rotation around the $y$-axis, with and angle $\theta = \arccos(e^{\kappa(t)})$.Here, we use as target the auxiliary qubit $Q_{\textrm{env}}$, and then a controlled-not gate using as target the system qubit $Q_{\textrm{sys}}$. At this point the state of $Q_{\textrm{sys}}$ corresponds to $\rho_{I}(t)$ in Eq.~(\ref{Amplitude_damping_map_1}), which correspond to the state at $t$, if we start with the state at time $0$. Third, the measurement step, where we measure $\hat{\sigma}_{x}$ or $\hat{\sigma}_{y}$, related to the expectation values of the memristor current and voltage as in Eq.~(\ref{2level_memristive_equation}). We note that the auxiliary qubit ($Q_{\textrm{env}}$) is initialized in the state $\ket{0}$.

Similarly, if we change the rotation angle $\theta$ in the evolution step, we can obtain the effective transformation from $t_i$ to $t_{i+1}$. To do this, we consider
\begin{equation}
    \kappa(t_{i+1},0) = \frac{{ - \int_0^{t_{i+1}} {\gamma (t')dt'} }}{2},\quad \kappa(t_i,0) = \frac{{ - \int_0^{t_i} {\gamma (t')dt'} }}{2} .
\end{equation}
Then, we can calculate $\hat{\rho}_I(t_{i+1})$ and $\hat{\rho}_I(t_{i})$ with Eq.~(\ref{Amplitude_damping_map_1}). Using the density matrices at time $t_i$ and $t_{i+1}$, we can find the dynamical map from $t_i$ to $t_{i+1}$ as
\begin{eqnarray}
\hat{\rho}_I(&&t_{i+1}) = \varepsilon_{t_{i+1},t_{i}}(\hat{\rho}_I(t_i))\nonumber\\
&&=\hat{E}_0^{(t_i, t_{i+1})} \hat{\rho}_I(0) (\hat{E}_0^{(t_i, t_{i+1})} )^\dagger + \hat{E}_1^{(t_i, t_{i+1})}\hat{\rho}_I(0) (\hat{E}_1^{(t_i, t_{i+1})})^\dagger,
\end{eqnarray}
where $\kappa(t_{i+1},t_{i}) = \frac{{ - \int_{t_{i}}^{t_{i+1}} {\gamma (t')dt'} }}{2}$, with $E_{0(1)}$ defined in Eq.~(\ref{Amplitude_damping_map_1}). Therefore, in our digital simulation, $\varepsilon_{t_{i+1},t_{i}}$ is the super operator for the memristive dynamics from $t_i$ to $t_{i+1}$.
\begin{figure}[t]
	\centering
	\includegraphics[width=1\linewidth]{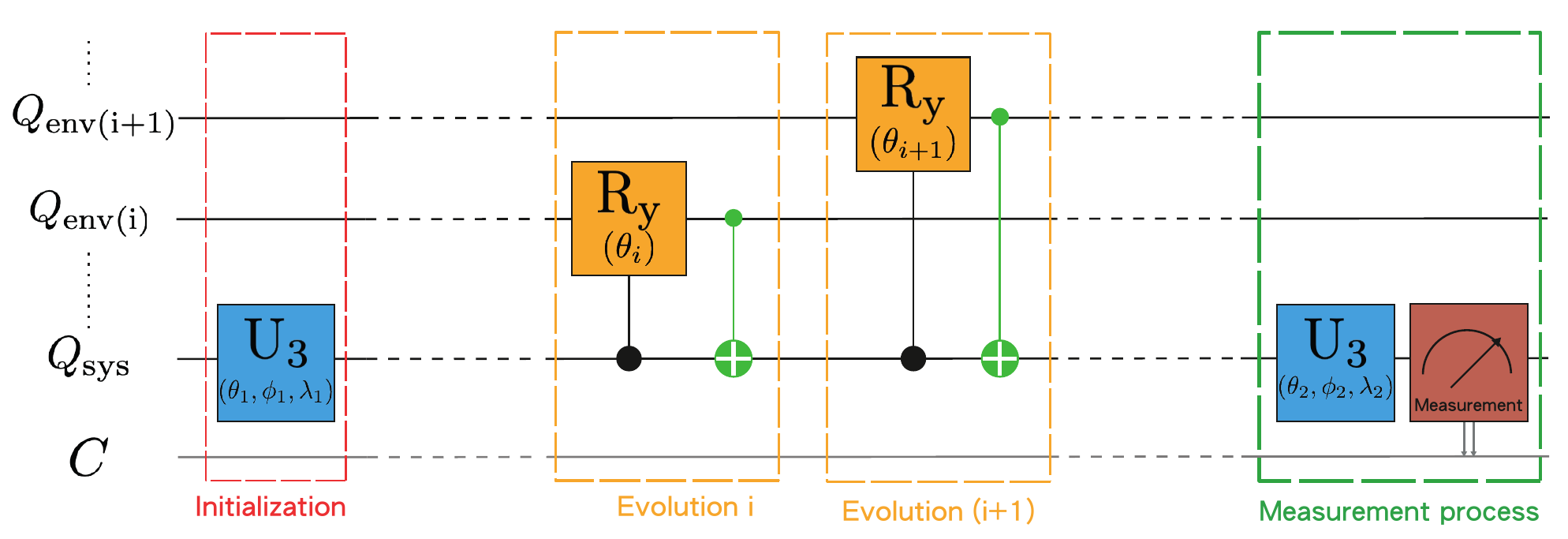}
	\caption{The proposed circuit for the digitized dynamics simulation. As an extension of the circuit shown in Fig.~\ref{Fig02}, the evolution step is repeated \emph{n} times before the measurement process.}
	\label{Fig03}
\end{figure}
The quantum circuit corresponding to this method is shown in Fig.~{\ref{Fig03}}, which is an extension of the circuit of Fig.~\ref{Fig02}. We require one ancillary qubit for each time step $\varepsilon_{t_{i+1},t_{i}}$ of the memristive qubit $Q_{\textrm{sys}}$. We can see that the rotation angle $\theta = \arccos[e^{\kappa(t_i,t_{i+1})}]$ only depends on the decay rate $\gamma(t)$ and the time steps. By measuring the expectation value of $\hat{\sigma}_x$ and $\hat{\sigma}_y$ after each step, we can obtain the evolution of the memristive variables in Eq.~(\ref{2level_memristive_equation}).

\begin{figure}[t]
	\centering
	\subfigure{\includegraphics[width=1.1\linewidth]{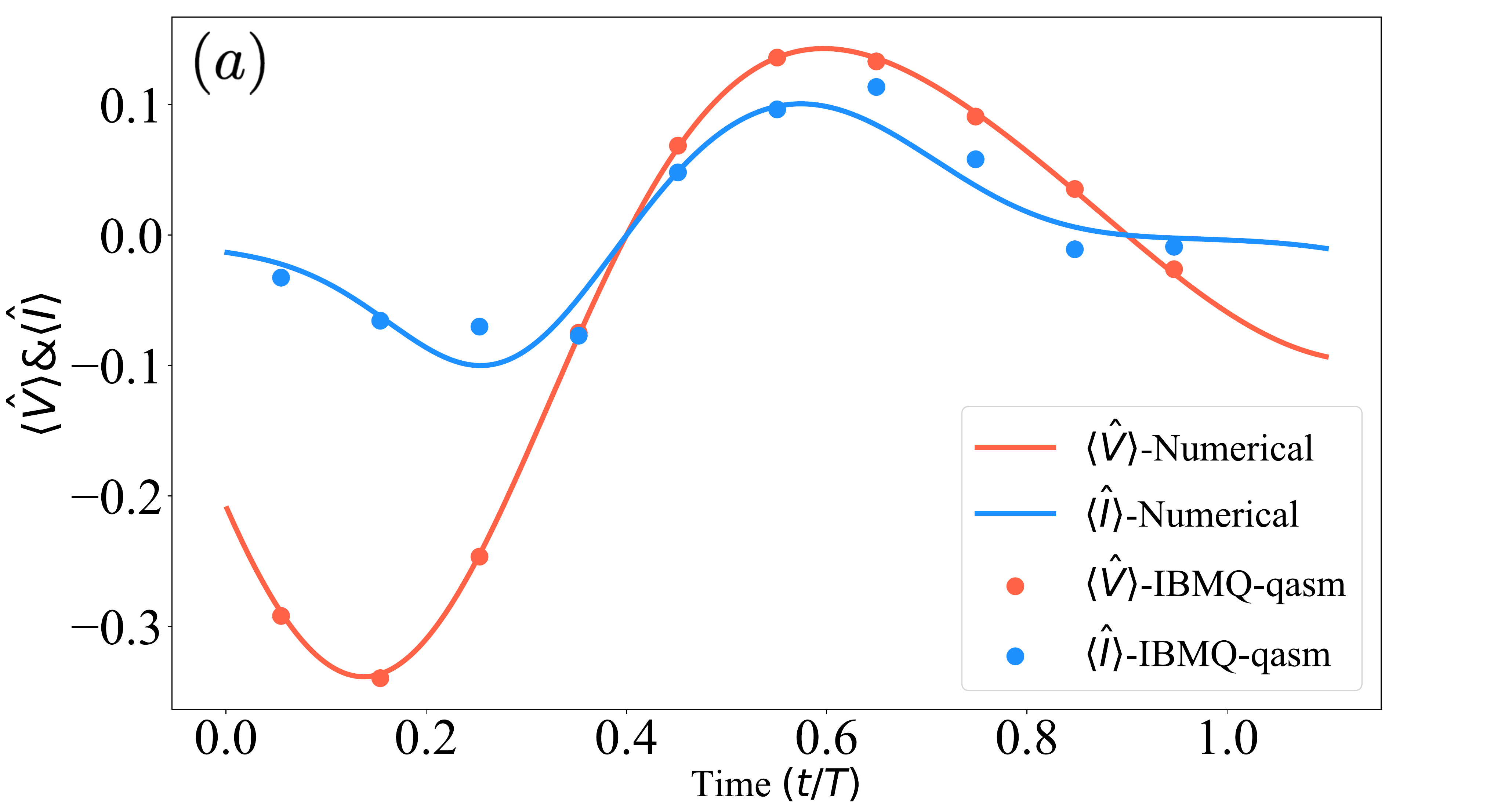}}
	\subfigure{\includegraphics[width=1.1\linewidth]{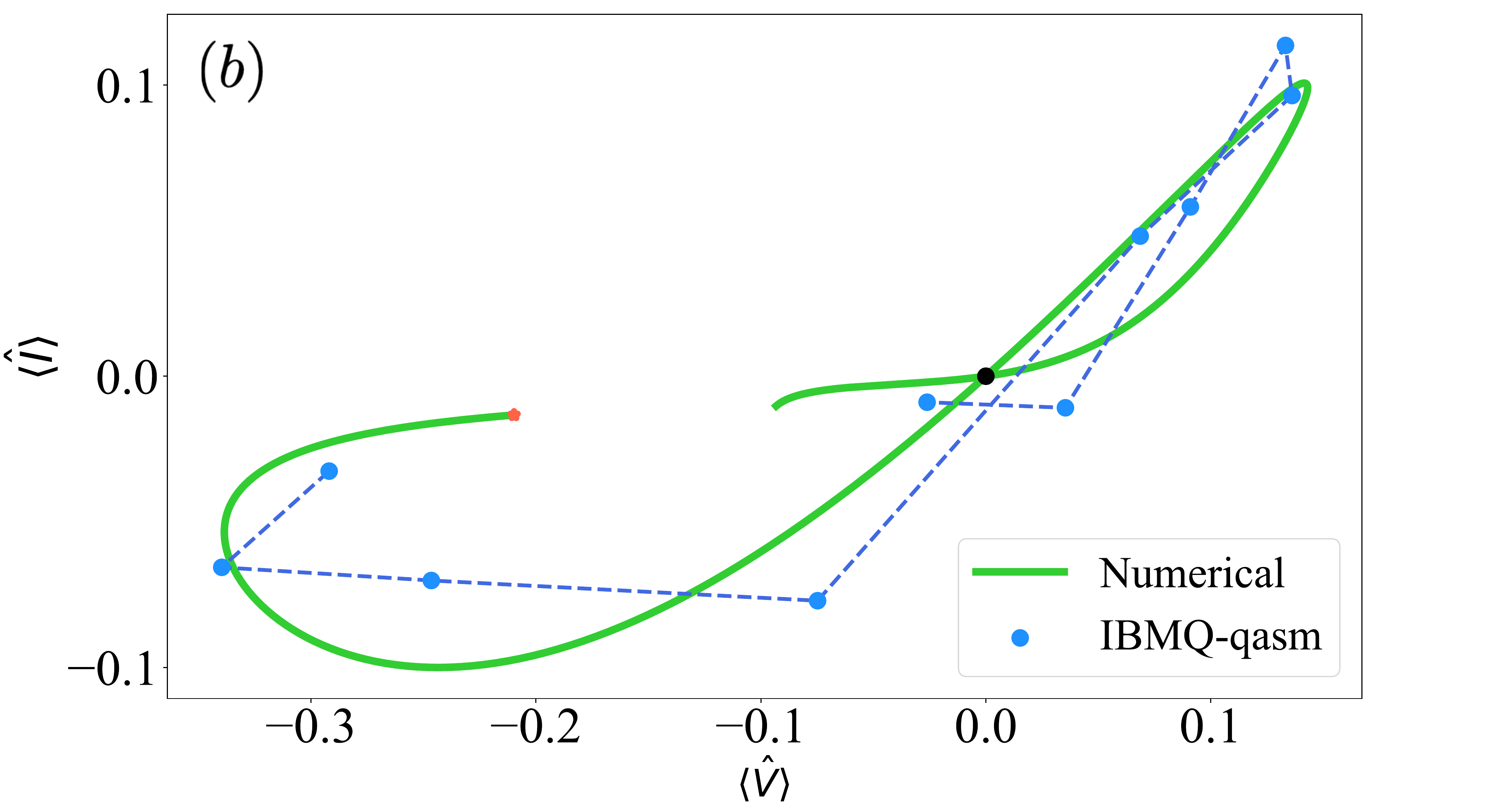}}
	\caption{(a)~Expectation values of the normalized current and voltage obtained from the digital (points) and numerical (lines) methods. The parameters are $m=1$, $\hbar=1$, $\omega=1$ and $\gamma_{0}=0.4$, and the initial state is ${\vert \psi_{0} \rangle = \cos(\pi/4)\ket{e} + \sin(\pi/4)e^{(i\pi/5)}\ket{g}}$. (b)~The I-V characteristics determined from the numerical (green line) and digital (blue points) methods. The red point denoted the starting point and the black point shows zero, the hallmark of a memristor's pinched hysteresis loop.}
	\label{Fig04}
\end{figure}

In Fig.~\ref{Fig04}, we show that the digital quantum simulation of a single quantum memristor compares well with the direct numerical solution of the master equation of Eq.~(\ref{mastereq2}). As can be seen, both the time evolution in Fig.~4a, as well as the memristor I-V characteristics in Fig.~4b, are well captured by our simulation protocol. Since the quantum memristor current calculation requires a time derivation, Eq.~(\ref{mem-eqs}), the simulation precision depends on the time resolution. In our calculations, we choose 30 points per period of oscillation. We note that the source of noise in the simulator is given by the statistical error due to the finite number of shots (measurements) to estimate the expectation value and the time derivatives. Specifically, we use $5000$ shots, which is the maximum allowed by the IBM quantum processor, obtaining a more realistic curve.

\vspace{0.5cm}

\subsection{Interaction between two memristors}~\label{subsec:3B}

So far, we have investigated a single quantum memristor dynamics simulation suitable for a digital quantum computer with a few qubits. However, the implementation of a neuromorphic quantum computer requires the coupling of many such memristive quantum devices. This field is a largely unexplored area and only a few efforts have been done to understand the entanglement in a network of coupled quantum memristors~\cite{Shubham2021PRA}. The digital quantum simulation of memristive quantum dynamics, as proposed here, allows one to explore many different types of coupling in a quantum computer. Below, we study several possible couplings between the quantum memristors utilizing the versatility provided by our proposed quantum simulations. 

\begin{figure}[t]
	\centering
	\includegraphics[width=1\linewidth]{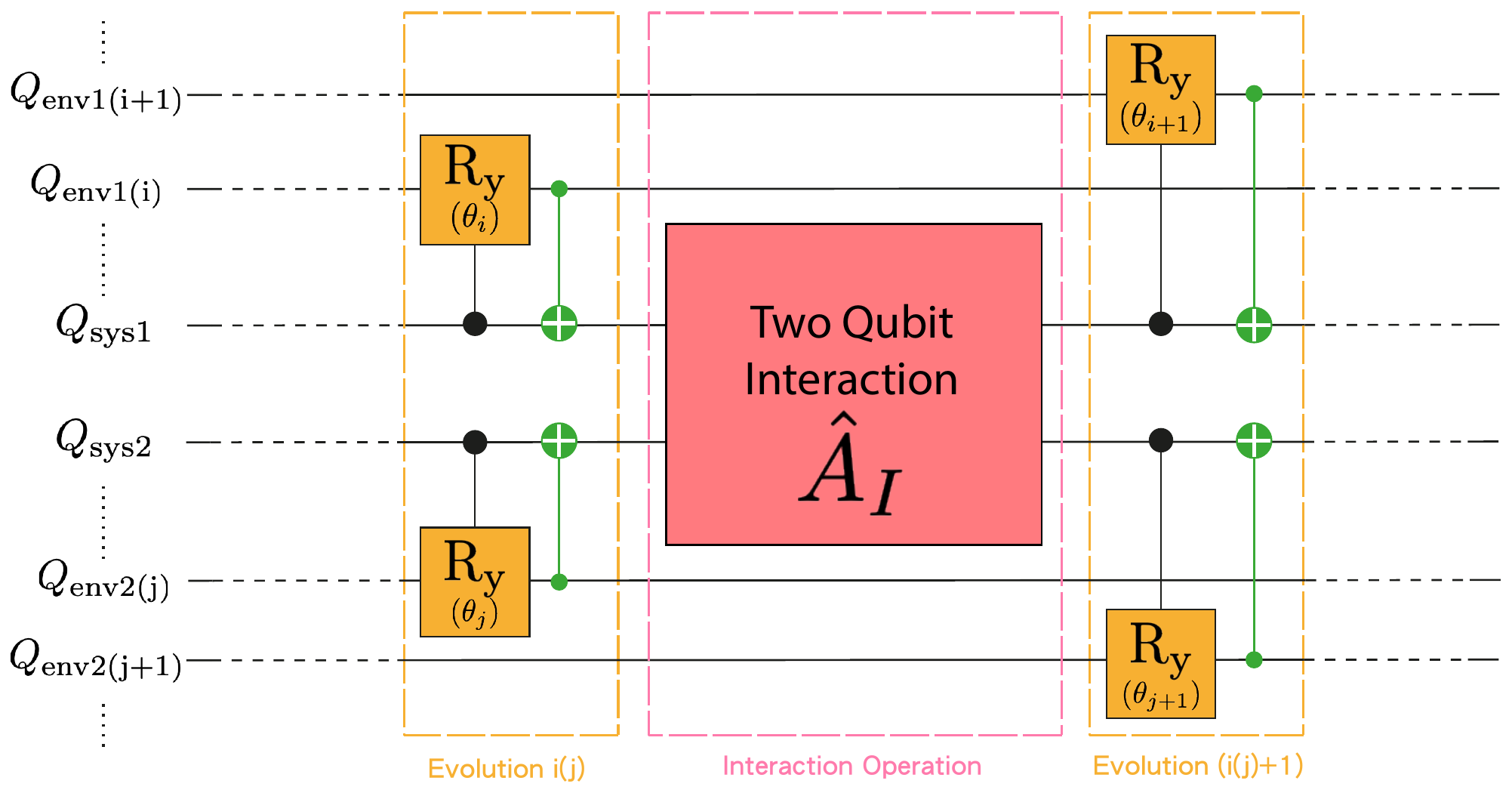}
	\caption{Coupling operation in the circuit. Each qubit, $Q_{\textrm{sys1}}$ and $Q_{\textrm{sys2}}$ undergo individual time evolution followed by their interaction, denoted by $\hat{A}$.}
	\label{Fig05}
\end{figure}

By increasing the qubit lines in the circuit of Fig.~\ref{Fig03}, we can simulate the dynamics of several quantum memristors and add interactions in each digital time step. This is shown in the circuit of Fig.~{\ref{Fig05}}, where the interaction operation $\hat{A}$ is applied at the end of each time steps $t_{i}$. By denoting the general expression of the two-qubit interaction as $\hat{A}$, we can write the evolution from time step $t_{i}$ to $t_{i+1}$ as
\begin{eqnarray}
    \hat{\rho}_{I}'(t_{i+1}) &=& \hat{A}^\dagger\hat{\rho}_I(t_{i+1})\hat{A},\nonumber \\
    \hat{\rho}_I(t_{i+1}) &=& \varepsilon_{1;t_{i},t_{i-1}}\otimes \varepsilon_{2;t_{i},t_{i-1}} \left(\hat{\rho}_{I}'(t_{i})\right),
\end{eqnarray}
where the subscript index in the dynamical map means that it only acts only on the subsystem 1(2) while the interaction gate $\hat{A}$ acts over both qubits, i.e. $Q_{\textrm{sys1}}$ and $Q_{\textrm{sys2}}$. 

Since coupling quantum memristors affects their individual hysteresis curves in non-trivial ways, we can characterize this effect by the form factor, $F$, which is defined as
\begin{equation}
    F = 4 \pi \frac{S}{P^2},
\end{equation}
which measures the form of a closed loop, where $S$ is the area enclosed by one loop in the hysteresis curve, and $P$ is its corresponding perimeter. We choose the form factor as it has been shown that the area enclosed by hysteresis curve is directly related to the memristor memory effects~\cite{Radwn2010IEEE,Biolek2012IEEE,Biolek2014EL}.

In what follows, we study two types of interaction; first, a native interaction given by the circuit implementation of the interaction Hamiltonian natural to this class of quantum memristors~\cite{Shubham2021PRA}. Second, a non-native interaction given by other combination of gates that are not representing the interactions that can be achieved in the architecture of superconducting quantum memristors. Here, we will discuss the cases which can preserve the memristive dynamics of each memristor, while several other cases are shown in appendix \ref{Ap}.

\begin{figure}[b]
	\centering
	\includegraphics[width=1\linewidth]{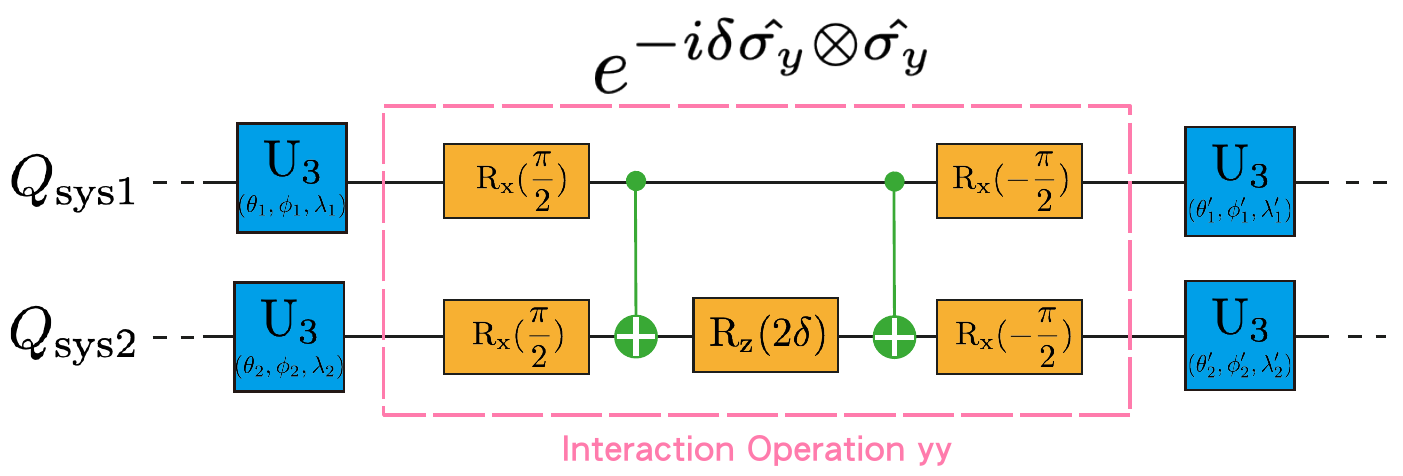}
	\caption{Implementation of a native interaction between two coupled quantum memristors in a digital quantum circuit.}
	\label{Fig06}
\end{figure}

{\it Native Interaction}.-- For this first class, we consider the gate decomposition of the unitary operation given by
\begin{equation}
    \hat U = e^{-i\delta \hat\sigma_{i}\otimes\hat\sigma_{i}} \, , \quad(i = x,y,z),
\end{equation}
which acts on the composite system, where $\delta$ is related to the coupling strength and the interaction time. In our digital circuit implementation, this unitary operation can be realized by adding a two-qubit gate between the two quantum memristive lines as shown in Fig.~\ref{Fig06}.

\begin{figure}[t]
	\centering
	\includegraphics[width=1\linewidth]{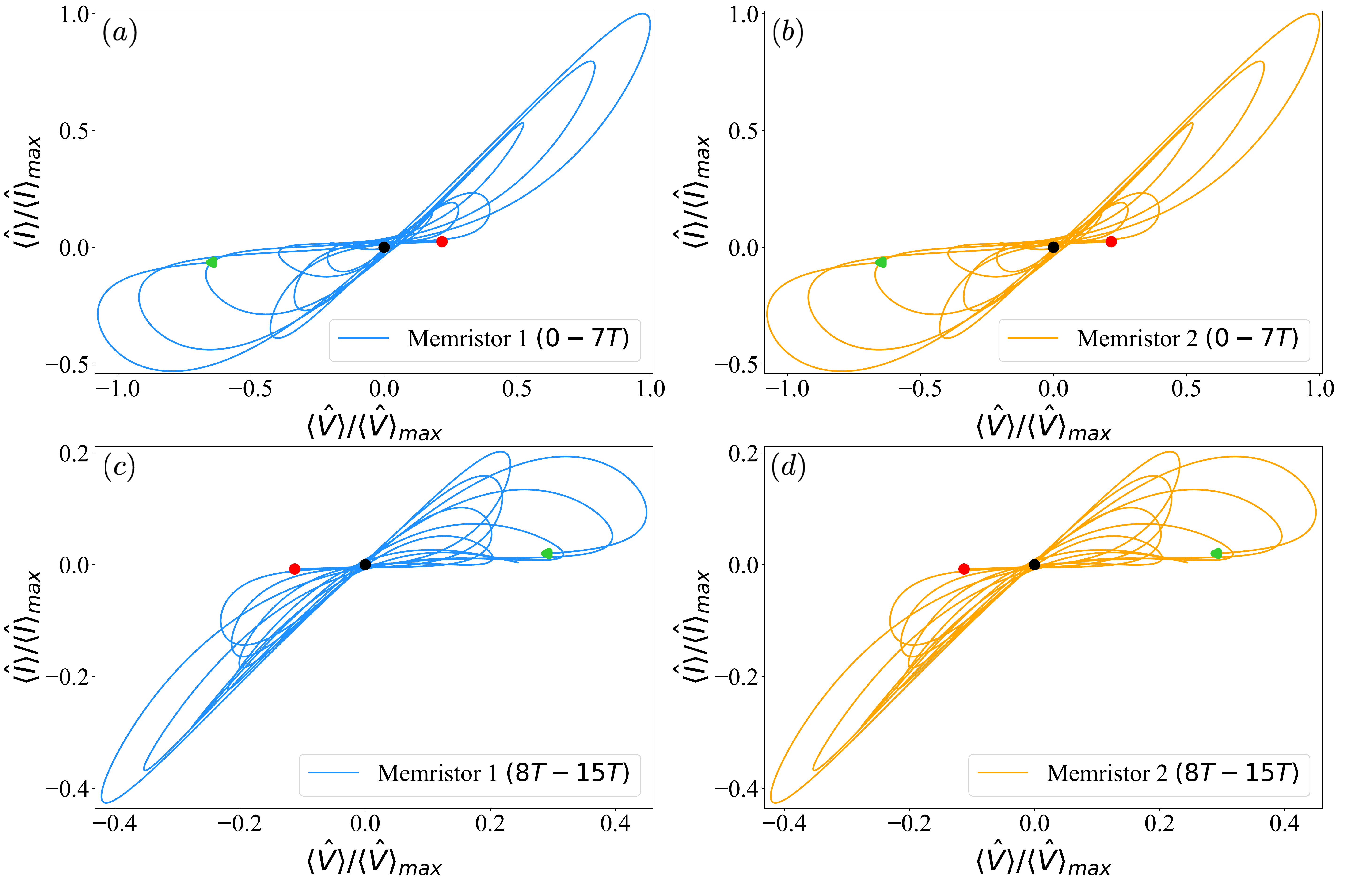}
	\caption{I-V curves for two coupled quantum memristors with $\hat{\sigma}_y \otimes \hat{\sigma}_y$ interaction. In each figure, the curve starts from the green point and ends at the red point. The black circle denotes the zero point where the hysteresis curves pinch. The expectation values of $\hat{I}$ and $\hat{V}$ are normalized by the maximum values. For both subsystems, we select $\ket{\psi_0} = {1/\sqrt{2}}(\ket{e}+\ket{g})$ as the initial state and they both have the same decay rates of $\gamma_0 = 0.02$. With the same parameters and symmetric interaction, both subsystems have the same dynamics and behavior of expectation value.}
	\label{Fig07}
\end{figure}

As an example, we consider the case of $\hat\sigma_{y}\otimes\hat\sigma_{y}$ interaction, which corresponds to the coupling between the internal variable in a asymetric SQUID quantum memristor. In Fig.~\ref{Fig07}, we show the I-V hysteresis plot for each subsystems, with identical initial states. For clarity, we plot the first 7 oscillations of the input, $\langle \hat{V} \rangle$, with period $T$ for memristor 1 and 2 in Fig.~\ref{Fig07} (a) and (b), respectively. In the same way, we plot from oscillation 8 to 15 in Fig.~\ref{Fig07} (c) and (d). We can see that the interaction changes the memristive behavior when compared to the uncoupled case, but each subsystem preserves their own pinched hysteresis curve hence, their memristive properties. 

\begin{figure}[ht]
	\centering
	\includegraphics[width=1\linewidth]{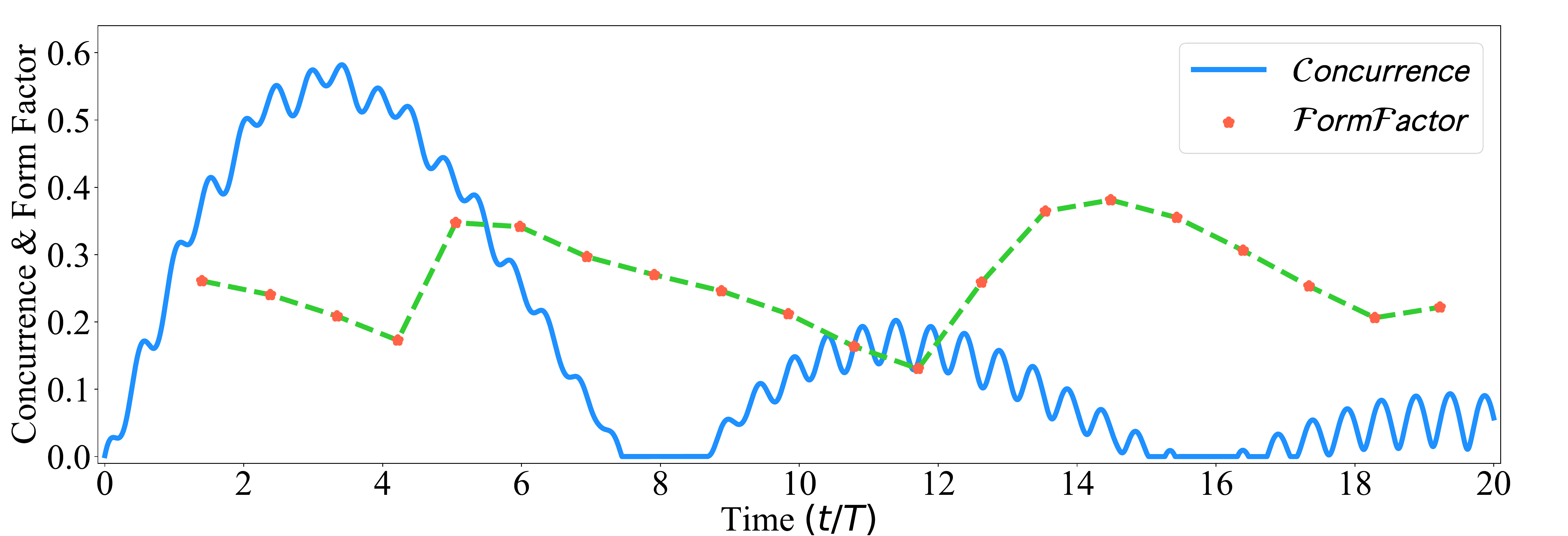}
	\caption{Concurrence and the form factor of the composite quantum  memristive system shown in Fig.~\ref{Fig07} as a function of time. During the 20 hysteretic loops, we obtain two significant peaks, one is from $t=0$ to $t=7T$ and the other is from $t=8T$ to $t=15T$.}
	\label{Fig08}
\end{figure}

In addition, we calculate the concurrence of the composite system and the form factor, shown in Fig.~(\ref{Fig08}). We can observe entanglement sudden death (ESD) and entanglement sudden birth (ESB) during the time evolution, and the behavior is inversely proportional to the form factor, which recovers the results obtained in Ref.~\cite{Shubham2021PRA}.

\begin{figure}[h]
	\centering
	\includegraphics[width=1.04\linewidth]{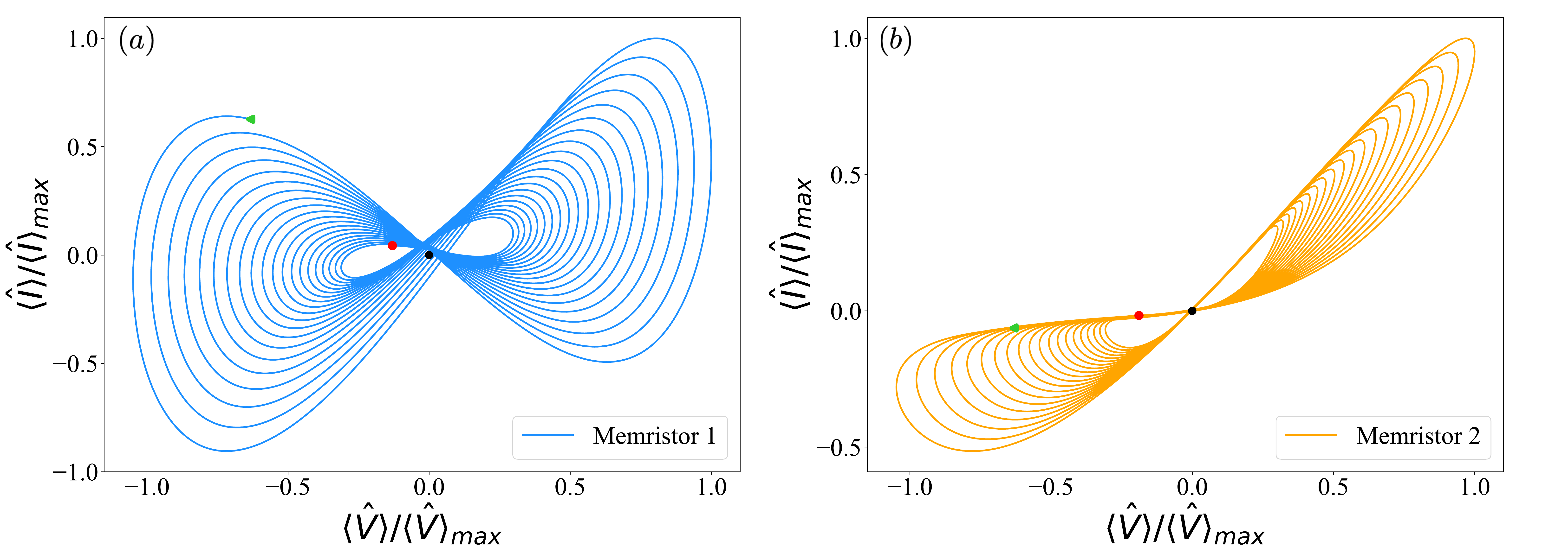}
	\caption{I-V curve for both subsystems with 20 oscillation periods. With the asymmetric interaction CRy, the two subsystems are changing differently. Both qubits starts from initial state $\ket{\psi_0} = {1/\sqrt{2}}(\ket{e}+\ket{g})$ and they have the same decay rate of $\gamma_0 = 0.02$. The dynamics of both start from the green triangle while the black circle is the zero point.}
	\label{Fig09}
\end{figure}

{\it Non-native Interaction}.-- Next, we analyze the non-native interaction between the two quantum memristors, which we consider as controlled two-qubit logical gates. The study of these cases is relevant since it can help to understand the mechanisms of manipulating the memristivity.

We consider interaction operators of the form
\begin{equation}
    \hat{C}_{Ri} = \ket{e}\bra{e}\otimes\hat{R_i}(\delta) + \ket{g}\bra{g}\otimes 1 \, ,\quad(i = x,y,z) \, .
\end{equation}
$\hat{R_i}(\delta)$ denotes rotation gates around a different axis corresponding to three different rotation directions in the Bloch sphere for a single qubit. 
For the case of controlled operations, one of the quantum memristors is chosen as control and the other as target system. The effect of the interaction between the two quantum memristors strongly depends on the state of the control quantum memristor.

As an example, we consider the controlled-Y gate as the interaction between the two subsystems. Figure~\ref{Fig09} shows the corresponding I-V curve for each quantum memristor for 20 oscillations. Figure~\ref{Fig09}a shows the I-V curve for the quantum memristor corresponding to the control qubit, and Fig.~\ref{Fig09}b shows the I-V curve for the target quantum memristor. With the implementation of such an asymmetric interaction, the resulting hysteresis curve is very different for each subsystems even when both have the same initial state. Here, the control memristive system has a bias in its loop for every period which means it does not satisfy the memristive criteria. In contrast, the subsystem 2 has successfully retained its memristive property. This can be understood as the target memristor perceiving the interaction with the control memristor as an additional collision with an environment, which can distort the hysteresis curve from its uncoupled form, but still retains its memristive character.

\begin{figure}[H]
	\centering
	\includegraphics[width=1\linewidth]{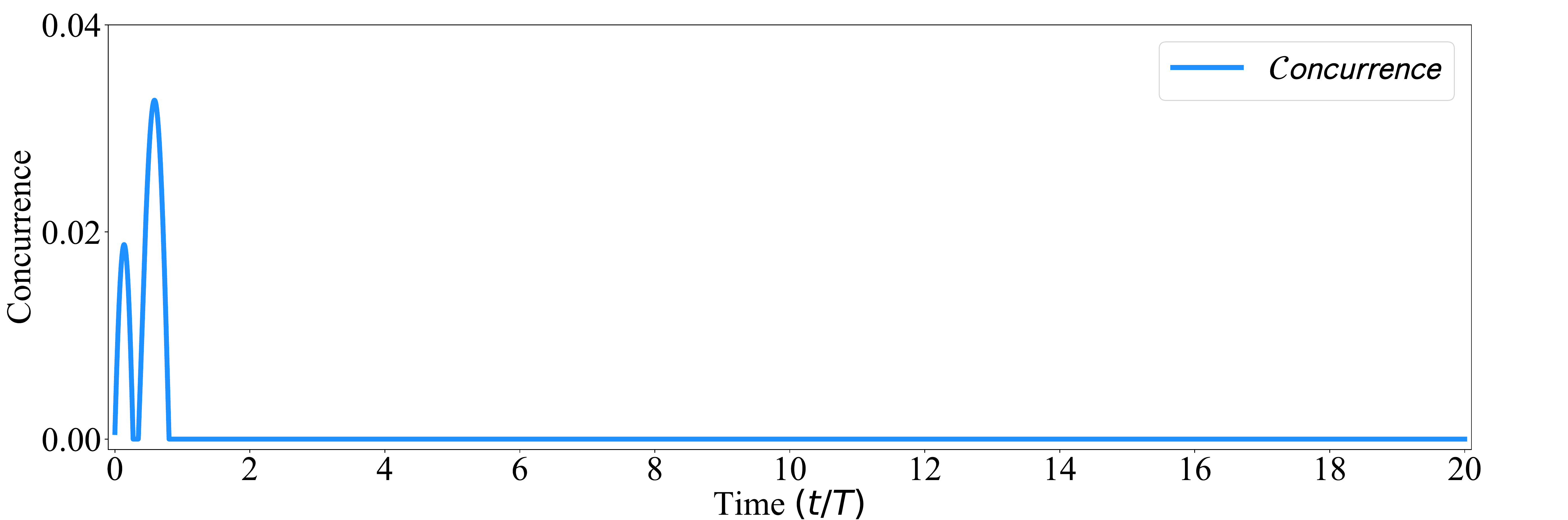}
	\caption{Concurrence of the composite quantum memristive system with CRy interaction. In total 20 oscillation periods, the concurrence only has some small increase within a short time interval.}
	\label{Fig10}
\end{figure}

Figure~\ref{Fig10} shows the concurrence of the composite system in time where we can see a small increase in the correlations during a short-time interval at the beginning. As time increases, the correlations between the subsystems quickly goes to zero, which can be associated with the underlying decay bringing the state of each subsystem to the ground state. 

\vspace{0.5cm}

\section{Conclusion}~\label{sec:4}

We have designed a quantum algorithm to simulate memristive quantum dynamics on a digital quantum computer. Our protocol employs a set of auxiliary qubits to simulate an effective environment which generates memristive quantum features in the expectation value of different operators in a single qubit. The memristive quantum dynamics has been obtained using each auxiliary qubit to evolve the system by one time step. We simulate our protocol using an IBM quantum simulator with 32 qubits and observe the hysteresis curve that characterizes a memristive quantum behavior. The statistical noise given by the total number of measurement shots, used to estimate the different expectation values, is the fundamental error in this class of quantum simulations. 

Furthermore, we consider interactions between the quantum memristors that can be achieved within the superconducting circuit architectures, and also interactions outside of this range. To analyze the effect of the interaction, we calculate the entanglement between the quantum memristors and the form factor for their corresponding hysteresis curves. We find that in the case of $\sigma_y \otimes \sigma_{y}$ interaction (native interaction) the memristivity is almost preserved, with a hysteresis curve that is slightly displaced from the origin. In the case of a  controlled-Y interaction (non-native interaction) between quantum memristors, the memristivity is perfectly preserved for the target quantum memristor and lost for the control quantum memristor. Other cases corresponding to different interactions result in complete loss of the memristive dynamics. In general, memristive quantum dynamics are fragile to interactions. In this sense, the cases that preserve the memristivity of each component can be of interest for future applications. Among them, we highlight the upcoming development of neuromorphic quantum computing for the design of quantum neural networks based on quantum memristors. In this context, neuromorphic quantum simulations provide a tool to explore coupled quantum memristive dynamics, not restricted to native interactions, in order to gain insight about the underlying mechanisms of quantum memristors. 

\vspace{0.5cm}

\centerline{\appendix{\bf APPENDIX}}

\section*{Coupled quantum memristors: other cases}\label{Ap}

\vspace{0.1cm}

We show further cases for the interaction of quantum memristors. Again, we have classified them under native interaction and non-native interaction.

\begin{figure}[H]
	\centering
	\includegraphics[width=1.05\linewidth]{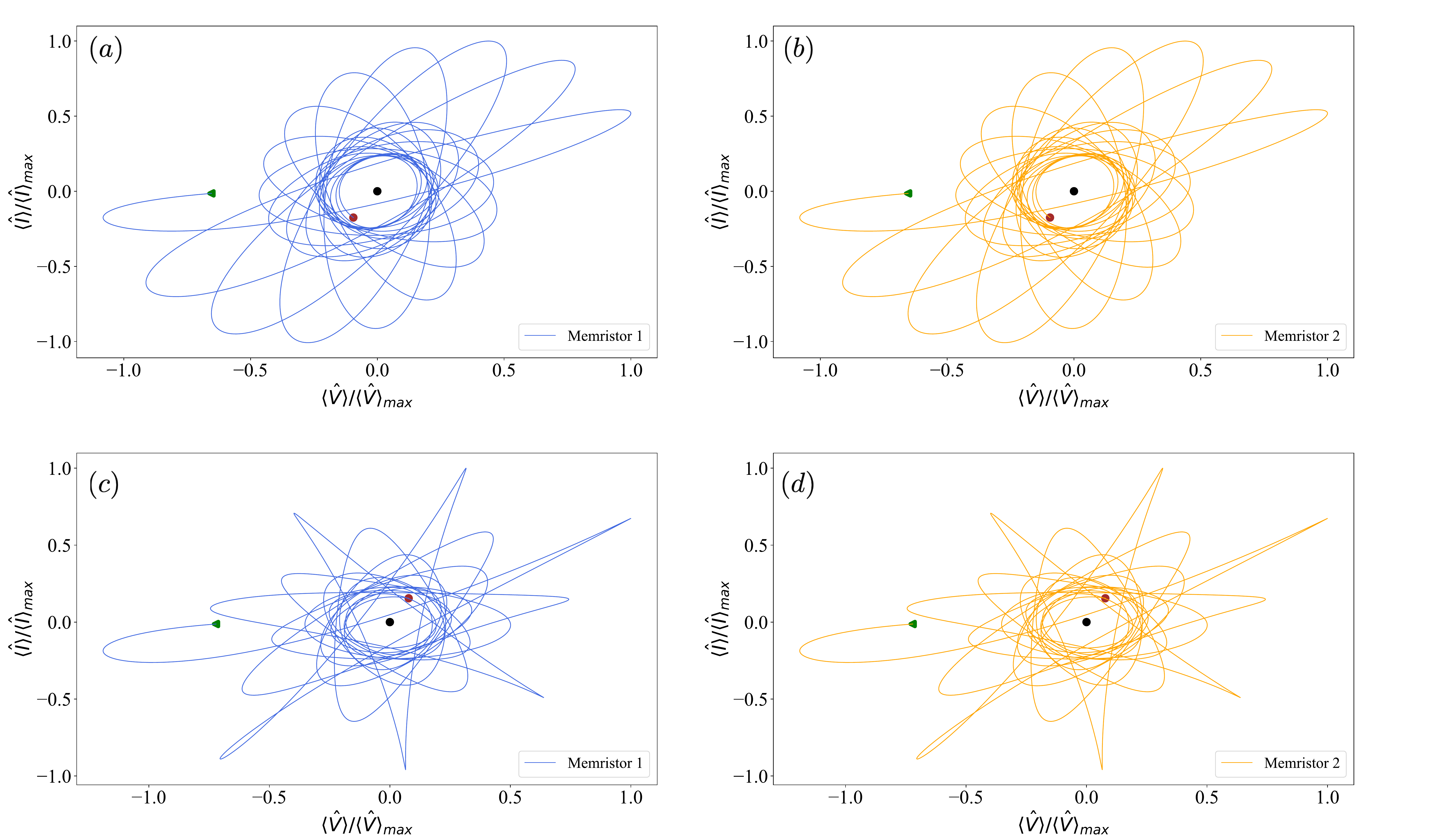}
	\caption{I-V curve of $\hat\sigma_{x}\otimes\hat\sigma_{x}$(a, b) and $\hat\sigma_{x}\otimes\hat\sigma_{x}$(c, d) interaction. We have considered 20 oscillations of $\langle \hat{V} \rangle$ for both cases. And we select $\ket{\psi_0} = {1/\sqrt{2}}(\ket{e}+\ket{g})$ as the initial state and they both have the same decay rates of $\gamma_0 = 0.02$.}
	\label{other_physical_inter}
\end{figure}

{\it Native interaction}.--  We consider the $\sigma_x \otimes \sigma_x$ and $\sigma_z \otimes \sigma_z$ interaction. Figures~\ref{other_physical_inter} (a) and (b) show the I-V curves for $\sigma_x \otimes \sigma_x$ interaction for each quantum memristor, and Figures~\ref{other_physical_inter} (c) and (d) show the I-V curves for $\sigma_z \otimes \sigma_z$ interaction. Notice that in these cases the interaction completely destroys the memristive dynamics.

\begin{figure}[H]
	\centering
	\includegraphics[width=1\linewidth]{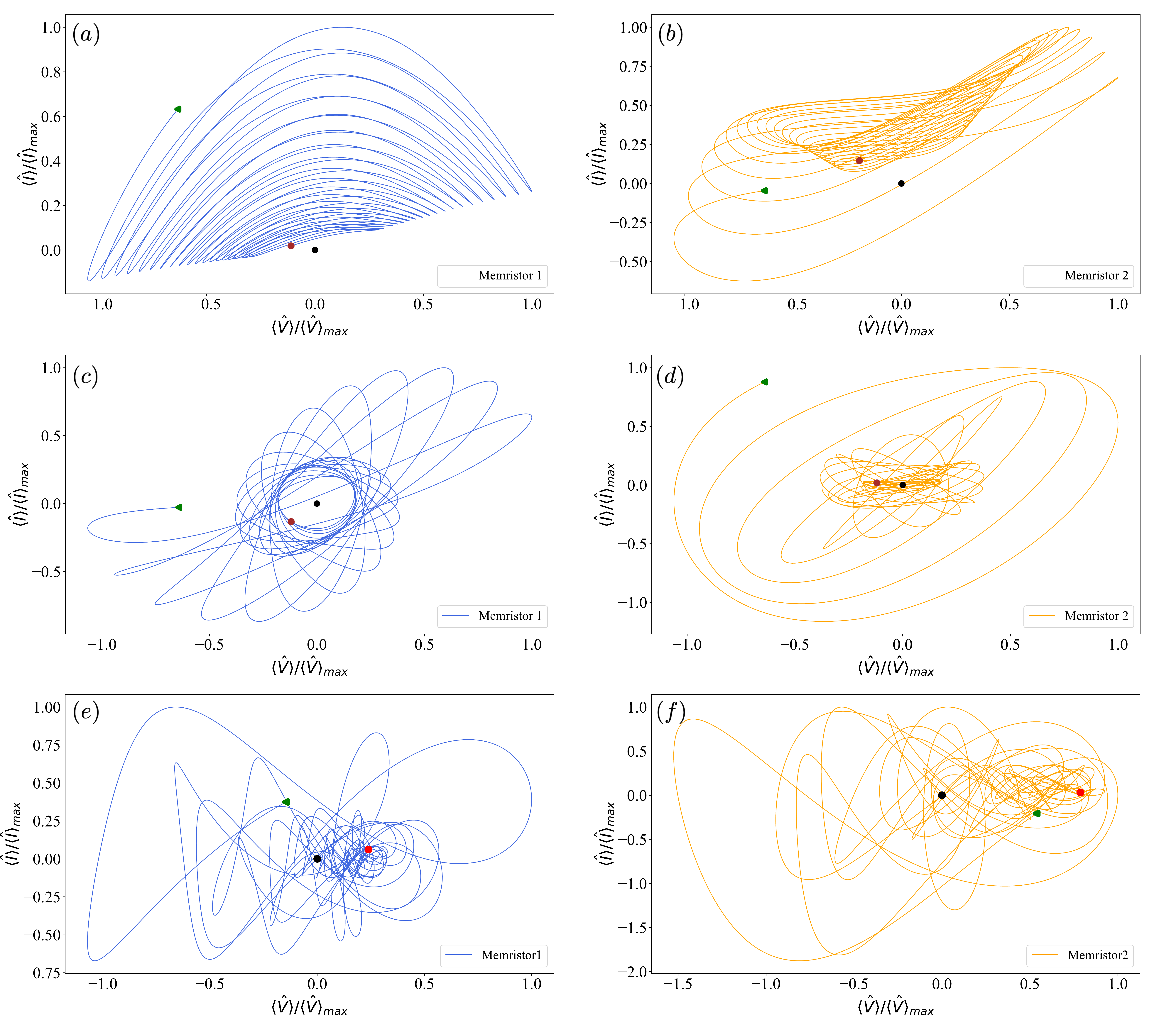}
	\caption{I-V curve of Controlled-Rotation-X(a, b), Controlled-Rotation-Y(c, d), and Partial-SWAP(e, f) interaction. We have considered 20 oscillations of $\langle \hat{V} \rangle$ for both cases. And we select $\ket{\psi_0} = {1/\sqrt{2}}(\ket{e}+\ket{g})$ as the initial state, where both have the same decay rates $\gamma_0 = 0.02$.}
	\label{other_nonphysical_inter}
\end{figure}

{\it Non-native interaction}.-- We consider the controlled-$X$, controlled$-Z$, and Partial-SWAP. The corresponding I-V curve for each quantum memristor is shown in Fig.~\ref{other_nonphysical_inter}. Again, in these cases the interaction completely destroys the memristive dynamics.

\clearpage

\bibliographystyle{apsrev}

\end{document}